\documentclass[review]{elsarticle}

\usepackage{lineno,hyperref}
\usepackage{color}
\usepackage{amsmath,amsfonts,amssymb,esvect,ulem}
\usepackage{float}
\usepackage{graphicx}
\usepackage{epsfig}
\usepackage{cancel}
\usepackage{epstopdf}
\usepackage{subcaption}
\modulolinenumbers[5]

\journal{Progress of Nuclear Energy}

\newcommand{\px}{\frac{\partial}{\partial x}}
\newcommand{\st}{\sigma_\mathrm{t}}
\newcommand{\pxt}{\frac{\partial^2}{\partial x^2}}

\newcommand{\psii}[1]{\psi^\mathrm{inc}_\mathrm{{#1}}}

\newcommand{\sn}{S$_N$}
\newcommand{\pn}{P$_N$}

\newcommand{\pd}{{\partial\mathcal{D}}}

\newcommand{\intli}[1]{\int\limits_{{#1}}}

\newcommand{\ndo}{\vec{n}\cdot\ome}
\newcommand{\absndo}{\left|\ndo\right|}

\newcommand{\e}[1]{\ensuremath{\times 10^{#1}}}

\newcommand{\sigs}{\sigma_\mathrm{s}}
\newcommand{\siga}{\sigma_\mathrm{a}}

\newcommand{\intd}{\int\limits_\mathcal{D}dV\ }
\newcommand{\intfpd}{\int\limits_{4\pi}d\Omega\ \int\limits_\mathcal{D}dV\ }

\newcommand{\ome}{\vv{\Omega}}
\newcommand{\dome}{d\Omega}

\newcommand{\lp}{\left(}
\newcommand{\rp}{\right)}

\newcommand{\bs}{\mathbf{S}}
\newcommand{\bl}{\mathbf{L}}
\newcommand{\ba}{\mathbf{A}}
\newcommand{\bst}{\Sigma_\mathrm{t}}

\newcommand{\half}{\frac{1}{2}}
\newcommand{\bmm}{\mathbf{M}^-}
\newcommand{\vq}{\vec{Q}}
\newcommand{\vphi}{\vec{\phi}}
\newcommand{\vt}{\vec{v}}
\newcommand{\qin}{q_\mathrm{inc}}

\newcommand{\icm}{cm$^{-1}$}

\newcommand{\quand}{\quad\mathrm{and}\quad}
\newcommand{\pp}[1]{P$_{#1}$}

\newcommand{\pord}[2]{\phi_{#1}^{(#2)}}

\begin{document}

\begin{frontmatter}

\title{Accurate Least-Squares P$_N$\ Scaling based on Problem Optical Thickness for solving Neutron Transport Problems}

\author[mymainaddress]{Weixiong Zheng\corref{mycorrespondingauthor}}
\ead{zwxne2010@gmail.com}

\author[ryanaddress]{Ryan G. McClarren}
\cortext[mycorrespondingauthor]{Corresponding author}
\ead{rgm@tamu.edu}
\address[mymainaddress]{Department of Nuclear Engineering, University of California, Berkeley, 4103 Etcheverry Hall, Berkeley, CA 94720}
\address[ryanaddress]{Texas A\&M Nuclear Engineering, 3133\ TAMU,~College Station, TX 77843-3133}
\begin{abstract}
	In this paper, we present an accurate and robust scaling operator based on material optical thickness (OT) for the least-squares spherical harmonics (LS\pn) method for solving neutron transport problems. LS\pn\ without proper scaling is known to be erroneous in highly scattering medium, if the optical thickness of the material is large. A previously presented scaling developed by Manteuffel, et al.\ does improve the accuracy of LS\pn\, in problems where the material is optically thick. With that method, however, essentially no scaling is applied in optically thin materials, which can lead to an erroneous solutions in the presence of a highly scattering medium. Another scaling approach, called the reciprocal-removal (RR) scaled LS\pn,\ which is equivalent to the self-adjoint angular flux (SAAF)\ equation, has numerical issues in highly-scattering materials due to a singular weighting.  We propose a scaling based on optical thickness that improves the solution in optically thick media while avoiding the singularity  in the SAAF formulation.
\end{abstract}

\begin{keyword}
	Least Square \pn,\ Neutron Transport Equation, Reactor Shielding, Optical Thickness, Scaling, Thick Diffusion Limit
\end{keyword}

\end{frontmatter}

\linenumbers

\section{Introduction}
\label{sect::intro}
The finite element method is widely used in solving partial differential equations (PDEs) due to features such as geometric flexibility and extensibility to higher-order basis functions. Yet, for first-order hyperbolic PDEs, such as the neutron transport equation, the continuous finite element method (CFEM) cannot easily be used due to stability issues. This goaded the development of the discontinuous Galerkin finite element method, which has  more degrees of freedom. In addition,  first-order systems can be difficult to solve using standard algebraic solvers {such as Krylov subspace methods} \cite{morel_saaf},\ motivating the development of second-order forms of transport equation and corresponding solution techniques.

The least-square finite element method is an attractive alternative method in solving neutron transport equations, which has been actively developed \cite{zheng-dissertation,zheng-l1,zheng_l1pn,zheng_l1sn,morel-holo,vincent-physor16,zheng-physor16} and implemented in production codes \cite{zheng-inl,yaqi-invite,clifmc}.\ It fixes the well-posedness issues of the standard Galerkin method and recovers some features of Ritz-Galerkin procedures, such as a symmetric positive definite (SPD) system, and thus enables the use of continuous finite element methods for the first-order neutron transport equation. To combine least-squares finite elements with the \pn\ method, however, the transport equation must be scaled before discretization in space and angle to obtain an accurate solution in thick diffusive regions \cite{manteuffel_lspn_scaling,manteuffel_boundary}. 

In order to improve the solution from LS\pn,\ Manteuffel, et al.\cite{manteuffel_lspn_scaling,manteuffel_boundary}\ developed a scaling operation for the LS\pn\ method that we refer to as the thick media (TM) scaling. Two shortcomings for this scaling have been observed. 
Firstly, as the scaling is originally presented, the scaling has different units for different equations, though this can be addressed by proper multiplication by a characteristic length scale. 
The second issue is that for materials in the ``thin"\ regime, i.e., $\st \ll 1$\ \icm, there is no scaling applied. Yet, in neutronics, especially shielding, detection, or thermal neutron scattering dominated problems \cite{zheng-freegas,zheng-emulation,Macf,thesis,physor,IKE,weixiong}\ such as light-water reactor applications, this can cause an issue when the optical thickness can be tens or hundreds of mean free paths (mfps.) although the macroscopic cross section is in the thin regime.

On the other hand, one could also scale the LS\pn\ by the  inverse of the removal operator. Such a scaling is equivalent to the self-adjoint angular flux (SAAF)-\pn\, equations discretized with CFEM. SAAF-\pn\, is known to be accurate in diffusive media. Nevertheless, in highly scattering medium with vanishing absorption, such a scaling is nearly singular, which leads to difficulty in preconditioning {when using algebraic solvers such as conjugate gradient method \cite{conjugate-gradient}}; the condition number of the resulting system can be large\cite{morel_saaf,clifmc}.

To overcome the shortcomings of those two scalings, we propose an optical thickness based scaling that scales the moment system based on mean free paths of the material with high scattering ratio, and limits to  SAAF-\pn, for other materials. On one hand, for thick scattering medium with small cross sections, such a scaling still scales the zeroth moment equation; on the other hand, our new scaling is not singular when $\sigma_\mathrm{a} \rightarrow 0$. As a result, the discretized system will not suffer from ill-conditioning in highly scattering media.

The outline of this paper is as follows. Section\ \ref{s:basics}\ introduces the basics of one group neutron transport in 1D slab geometry with a least-squares discretization in space and spherical harmonics\footnote{In 1D slab geometry, orthonormal Legendre polynomials, which is a special case of spherical harmonics, will be used instead.}\ in angle; Section\ \ref{s:scaling}\ introduces several scalings compared in this work; Section\ \ref{s:num}\ presents numerical tests for the proposed method; Section\ \ref{s:con}\ concludes this study.

%
\section{One-Speed Neutron Transport Equation and Least Square Functionals}
\label{s:basics}
\subsection{Transport equation with \pn\ approximation in angle}
The one-speed neutron steady state transport equation with isotropic scattering for direction $\ome$\ can be expressed as \cite{morel_saaf}
\begin{subequations}\label{e:te}
	\begin{equation}
	\vec{\Omega}\cdot\nabla\psi(\vec{r},\vec{\Omega})+\sigma_\mathrm{t}\psi(\vec{r},\vec{\Omega})={S}\psi(\vec{r},\vec{\Omega})+q(\vec{r},\vec{\Omega}),
	\end{equation}
	\begin{equation}
	\psi(\vec{r},\ome)=\psii{},\quad\vec{r}\in\pd\quand\ndo<0
	\end{equation}
\end{subequations}
where  $\psi$\ is the angular flux, $\vec{n}$\ is the outward normal vector on the boundary $\pd$\ and $\psii{}$\ is the incident flux. We have written the scattering source, ${S}\psi(\vec{r},\vec{\Omega})$, scalar flux, $\phi(\vec{r})$, and isotropic source $Q$, as
\begin{equation}
{S}\psi(\vec{r},\vec{\Omega})=\frac{\sigs\phi(\vec{r})}{4\pi},\quad\phi(\vec{r})=\intli{4\pi}\dome\ \psi(\vec{r},\ome)\quad\mathrm{and}\quad q(\vec{r},\vec{\Omega})=\frac{Q}{4\pi}.
\end{equation}
It will be convenient to express Eq.~\eqref{e:te} in operator form as 
\begin{subequations}\label{e:te2}
	\begin{equation}
	{(L-S)}\psi=q,
	\end{equation}
	with
	\begin{equation}\label{e:ldf}
	{L}=\vec{\Omega}\cdot\nabla(\cdot)+\sigma_\mathrm{t}.
	\end{equation}
\end{subequations}

In this work, we will treat the angular variable, $\Omega$, with the \pn\ method, i.e., the angular flux is expressed in terms of a truncated spherical harmonics expansion with the spherical harmonics written as $\left\{R_l^m(\ome)\right\}_l^m,\ l\leq N, |m|\leq l$.\ The spherical harmonics are normalized such that:
\begin{equation}
\intli{4\pi}\dome\ R_l^m(\ome)R_{l'}^{m'}(\ome)=\delta_{ll'}\delta_{mm'},
\end{equation}
where $\delta$\ is the Kronecker $\delta$\ function. The \pn\ method is derived using a Galerkin procedure with the spherical harmonics as test and weight functions to obtain a system of equations with expansion coefficients $\phi_l^m(\vec{r})$,\ also called moments, as the unknowns\cite{weixiong_tpn,vincent_fpn,mccfpn09}.\ That is, the transport equation is approximated as the system:
\begin{subequations}
	\begin{equation}
	(\bl-\bs)\vec{\phi}=\vec{Q}
	\end{equation}
	\begin{align}
	\bl=\nabla\cdot\ba+\bst,\quad\ba=\sum_{\zeta=x,y,z}\mathbf{e}_\zeta\intli{4\pi}\dome\ \Omega_\zeta\vec{R}(\ome)\vec{R}^\top(\ome)\nonumber\\
	\bst=\mathrm{diag}\{\st,\cdots,\st\}\quand\vec{Q}=\left(\frac{Q}{\sqrt{4\pi}},0,\cdots,0\right)^\top
	\end{align}
	\begin{align}
	\bs=\mathrm{diag}\{\sigs,0,\cdots,0\},
	\end{align}
\end{subequations}
where $\vec{R}$\ is a column vector containing all spherical harmonics functions up to Degree $N$.
\subsection{Least-squares functional of residuals and weak formulations}
\label{s:ls0}
Least-squares finite element methods for the first-order PDE $\mathcal{L}u=f$\ start by defining a quadratic functional in some finite element space $\mathcal{V}$
\begin{equation}\label{e:func0}
\mathcal{J}=\frac{1}{2}\intd\left(\mathcal{L}u-f\right)^2,
\end{equation}
which leads to the following minimization problem:
\begin{equation}
u=\operatornamewithlimits{min}_{u}\mathcal{J},\quad u\in\mathcal{V}.
\end{equation}
This results in the weak formulation in the spatial domain $\mathcal{D}$\ as follows\footnote{See Ref. \cite{runchang}\ for the procedure of deriving weak formulation from a least-square functional.} with any test function $v\in\mathcal{V}$:
\begin{equation}
\intd\mathcal{L}v\mathcal{L}u=\intd\mathcal{L}vf
\end{equation}

Using the least-squares method to solve transport problems is similar to the procedure illustrated above. Define a least-squares functional in space-angle as the following:
\begin{equation}\label{e:ls_func}
\mathcal{J}_0=\half\intfpd\lp(L-S)\psi-q\rp^2+\half\intli{\ndo<0}\dome\ \intli{\pd}ds\ \absndo(\psi-\psii{})^2,
\end{equation}
where $\psi(\vec{r},\ome)=\sum\limits_{l<N}\sum\limits_{|m|<l}R_l^m(\ome)\phi_l^m(\vec{r})$\ with $\phi_l^m(\vec{r})$\ the to-be-approximated moments spanned by user-defined finite element space. The term in this equation involving the half-range integral accounts for the boundary conditions.

Alternatively, for orthonormal spherical harmonics, Varin\cite{varin_dissertation}\ provided an equivalent way of deriving the least-squares method through defining a least-squares functional for \pn\ system to replace the interior functional in Eq.\ \eqref{e:ls_func}:
\begin{equation}\label{e:ls_func1}
\mathcal{J}=\half\intd\left(\left(\bl-\bs\right)\vphi-\vq\right)^2+\half\intli{\ndo<0}\dome\ \intli{\pd}ds\ \absndo(\psi-\psii{})^2,
\end{equation}

The resulting weak form is:
\begin{align}\label{e:ls_weak1}
\intd\left((\bl-\bs)\vec{v}\right)^\top(\bl-\bs)\vec{\phi}+\intli{\pd}ds\ \vt^\top\bmm\vphi\nonumber\\
=\intd\left((\bl-\bs)\vec{v}\right)^\top\vq+\intli{\pd}ds\ \vt^\top\qin,
\end{align}
where
\begin{align}
\bmm=\intli{\ndo<0}\dome\ \absndo\vec{R}\vec{R}^\top\quand\qin=\intli{\ndo<0}\dome\ \absndo\vec{R}(\ome)\psi^\mathrm{inc}(\ome),
\end{align}
which is a formulation of the \pn\ equations with least-squares finite element method.

Eqs.\ \eqref{e:ls_func}\ and \eqref{e:ls_func1}\ are very similar.\ In the remainder of this paper, however, we will utilize the latter formulation, i.e., defining a least-squares functional for the \pn\ system. The motivation is such that all types of scalings are then transformed to multipliers of different moment equations\cite{varin_dissertation,ressel_dissertation}\ directly applicable to the weak forms instead of operators for angular flux.
\subsection{Reciprocal removal scaled least-squares functional and weak formulations}
The drawback of solving the transport equation using methods in Section\ \ref{s:ls0}\ is that in highly scattering media with moderate optical thickness, the solution is erroneous {due to the fact that it does not preserve the asymptotic diffusion limit as discussed in \cite{manteuffel_lspn_scaling,varin_lspn}}.\ In order to alleviate the solution in such a situation, one could define a scaling operator when defining the functional using the inverse of removal operator, i.e.
\begin{equation}
\mathcal{J}_\mathrm{R}=\half\intd(\bst-\bs)^{-1}\lp(\bl-\bs)\vphi-{\vec{Q}}\rp^2+\half\intli{\ndo<0}\dome\ \intli{\pd}ds\ \absndo(\psi-\psii{})^2.
\end{equation}
From this functional, one can derive the following weak-form:
%
\begin{align}\label{e:saaf2}
\intd\left((\bl-\bs)\vec{v}\right)^\top(\bst-\bs)^{-1}(\bl-\bs)\vec{\phi}+\intli{\pd}ds\ \vt^\top\bmm\vphi\nonumber\\
=\intd\left((\bl-\bs)\vec{v}\right)^\top(\bst-\bs)^{-1}\vq+\intli{\pd}ds\ \vt^\top\qin.
\end{align}
{This being said, LS  scaled by the reciprocal removal operator} is equivalent to the weak form of the SAAF-\pn\ equations solved with CFEM.
\section{Other scalings}\label{s:scaling}
The operator $(\st-S)^{-1}$\ is singular and ill-posed for a pure scatterer \cite{morel_saaf}.\ As the scattering ratio approaches one, this operator is numerically ill-posed and can cause other issues,  especially in multi-D applications, including degradation of preconditioning\cite{clifmc}. On the other hand, without a scaling, the LS method can be easily error-prone for optically thick problems\cite{manteuffel_lspn_scaling,manteuffel_boundary}. To overcome these issues, Manteuffel, et al.\ proposed a scaling using the following constants
\begin{equation}
\tau_0=\begin{cases}
1.0,&\st<1\ \mathrm{cm}^{-1}\\
\st,&\st>1\ \mathrm{cm}^{-1}, \quad \text{and} \quad \st>1/\siga\\
1.0/\siga,&\mathrm{otherwise}
\end{cases},
\end{equation}
\begin{equation}
\tau_1=\begin{cases}
1.0,&\st<1\ \mathrm{cm}^{-1}\\
1/\st,&\mathrm{otherwise}
\end{cases},
\end{equation}
and the scaling matrix $\Lambda$\ as the matrix representation of the corresponding scaling operator\cite{ressel_dissertation}:
\begin{equation}
\Lambda=
\begin{bmatrix}
\tau_0& & &\\
&\tau_1&&\\
&&\ddots&\\
&&&\tau_1
\end{bmatrix}
\end{equation}
and the scaled least-squares formulation is:
\begin{align}
\intd\left((\bl-\bs)\vec{v}\right)^\top\Lambda(\bl-\bs)\vec{\phi}+\intli{\pd}ds\ \vt^\top\bmm\vphi\nonumber\\
=\intd\left((\bl-\bs)\vec{v}\right)^\top\Lambda\vq+\intli{\pd}ds\ \vt^\top\qin.
\end{align}
We will call this scaling the thick media scaling.

\subsection{An optical-thickness-based scaling}
The least-squares method without scaling with optically-thick spatial cells  does not preserve the asymptotic diffusion limit of neutron transport. The thick media scaling aims to improve the solution  by increasing the scaling of the zeroth moment  and decreasing the scaling of the  other moments.

There are, however, several potential drawbacks of the thick media scaling. In {many cases}, material cross sections are in the thin regime, i.e. $\st<1.0$\ \icm.\ Nonetheless, it does not necessarily mean the material is optically thin to neutrons, e.g. several meters of graphite shield. In fact, in situations such as shielding and detector problems, neutrons may need to pass through tens even hundreds of mean free paths to approach targets/detectors. If the material has a high scattering ratio, but in thin regime, the least-squares method can fail without a scaling. There are not many materials that {have cross sections falling} in thick regime with high scattering ratio, so in neutronics the thick medium scaling is of  limited  usefulness despite its theoretical value in applying scaling to reduce simulation errors.

On the other hand, for the limited numbers of scattering media with $\st>1$\ cm, like water, nevertheless, $\tau_0=\st$\ is fairly close to the value of $\tau_1=1/\st$.\ Moreover, in media such as graphite and heavy water, for which $\st<1$\ \icm\  there is a small amount of scaling.

Simultaneously, the SAAF formulation, or equivalently the RR scaled least-squares, $1/\siga$\ can be extremely large\footnote{For instance, heavy water has a macroscopic absorption cross section $3.3\e{-5}$\ \icm\ for room temperature (RT) neutrons and consequently $1/\siga>1.0\e{4}$\ cm \cite{xie_book}.}, which would potentially cause problems in multi-D calculations for heterogeneous problems.

Then we propose a scaling based on the optical thickness (OT). We also give different scaling factors $\tau_0$\ and $\tau_1$\ to the zeroth and other moment equations, respectively. The scaling satisfies the following requirements:
\begin{itemize}
	\item $\tau_0$\ should have the same units as $\tau_1$;
	\item $\tau_0\geq\tau_1$;
	\item $\tau_0$\ is bounded in optically thick shielding problems with highly scattering medium.
\end{itemize}
We therefore propose the following scaling:
\begin{equation}\label{e:tau}
\tau_0=\begin{cases}
\st DL,&1/\siga>\st DL>1/\st, \quad \text{and} \quad \siga<0.1\st\\
1/\siga,&\mathrm{otherwise}
\end{cases}
\end{equation}
\begin{equation}
\tau_1=1/\st,
\end{equation}
where $D$\ is a measure of the size of the medium and $L$\ is a characteristic length. For the tests in this study we use  $L=\max(1.0\ \mathrm{cm}, h)$. Overall, the scaling considered two factors affecting the scaling effectiveness, i.e., the medium optical thickness and local cell size. 
The formulation is such that in an optically-thick, scattering medium it gives a bounded, but non-zero, scaling to the zeroth moment equation.

\subsection{Asymptotics of the scaling}\label{s:asym}
The standard asymptotic diffusion limit analysis\cite{larsen_1987,spn_derive}\ makes the following scaling with $\epsilon >0$:
\begin{equation}\label{e:pscale}
\siga\to\epsilon\siga,\quad\st\to\frac{1}{\epsilon}\st\quand q\to\epsilon q.
\end{equation}
A scheme is said to preserve the thick diffusion limit of the transport equation if it limits to a discretization of the diffusion equation as $\epsilon \rightarrow 0$ \cite{larsen_1987,larsen_1989}.\ Using SAAF with CFEM preserves the thick diffusion limit. Therefore, given the equivalence of the two forms, solving the RR-scaled least-square equations with continuous basis functions, c.f.\ Eq.\ \eqref{e:saaf2}, will preserve the thick diffusion limit. 

To compare the scaling that yields SAAF, and our new scaling we will compare the two under the asymptotic scaling. Making the substitutions in Eq.~\eqref{e:pscale} to the RR scaling gives
\begin{equation}
\tau_0=\frac{1}{\siga}\to\frac{1}{\epsilon\siga}\sim O\lp\frac{1}{\epsilon}\rp\quand\tau_1=\frac{1}{\st}\to\frac{\epsilon}{\st}\sim O\lp\epsilon\rp.
\end{equation}
In the same manner, given that $DL = O(1)$, the OT scaling becomes
\begin{equation}
\tau_0=\st DL\to\frac{1}{\epsilon}\st DL\sim O\lp\frac{1}{\epsilon}\rp\quand\tau_1=\frac{1}{\st}\to\frac{\epsilon}{\st}\sim O\lp\epsilon\rp.
\end{equation}

The OT scaling has the same orders as the RR scaling for different moment equations. Therefore, the scaling is expected to have a similar asymptotic behavior as RR scaling, or equivalently SAAF. We demonstrate this using 1-D slab geometry away from boundaries. For a discussion of the effect of the boundary terms, see Varin, et. al \cite{varin_lspn}.

In 1-D slab geometry, the OT scaling leads to the following relationships for the first few orders of $\epsilon$:
\begin{subequations}
	\begin{align}
	\pord{1}{0}=0
	\end{align}
	\begin{align}\label{e:dif0}
	b\st DL\px R_\mathrm{D}+\st(\siga\st DL-1)\left(\pord{1}{1}+\frac{b}{\st}\px\pord{0}{0}\right)=0,
	\end{align}
	\begin{align}
	R_\mathrm{D}=\frac{-b^2}{\st}\pxt\pord{0}{0}+\siga\pord{0}{0}-Q_0,
	\end{align}
\end{subequations}
where $b=\frac{1}{\sqrt{3}}$\ is the normalization factor\footnote{Note that since we use orthonormal Legendre polynomial, $\phi_0=\phi/\sqrt{2}$\ and $Q_0=Q/\sqrt{2}$}. In these equations, the values of $\sigma_\mathrm{a}$ and $\sigma_\mathrm{t}$ are independent. Therefore, the two terms in  Eq.\ \eqref{e:dif0} must hold individually:
\begin{subequations}
	\begin{align}\label{e:dif1}
	\px R_\mathrm{D}=\px\left(\frac{-b^2}{\st}\pxt\pord{0}{0}+\siga\pord{0}{0}-Q_0\right)= 0,
	\end{align}
	\begin{align}\label{e:fick}
	\pord{1}{1}+\frac{b}{\st}\px\pord{0}{0}= 0.
	\end{align}
\end{subequations}

Equation \eqref{e:fick} is a version of the Fick's law between $\pord{1}{1}$\ and $\pord{0}{0}$.\ 
Equation \eqref{e:dif1}\ is the  correct diffusion equation. Therefore, we expect the OT scaling to provide accurate solutions in the thick diffusion limit.  We will show this in numerical results next.

%
\section{Numerical Tests}
\label{s:num}
The numerical tests are carried out with the {\tt C{\small ++}} open source finite element library deal.II\cite{dealii82}. {Several tests within the optically thick regime are conducted. As the OT-LS\pn\ resembles CFEM-SAAF-\pn,\ which has been studied in the literature in the thin region, we do not produce optically thin results here. Interested readers can refer to \cite{clifmc,cao-saaf-pn,saaf-thesis}.}
\subsection{Asymptotic test}
\begin{figure}[ht!]
	
	\centering
	\includegraphics[width=.75\linewidth]{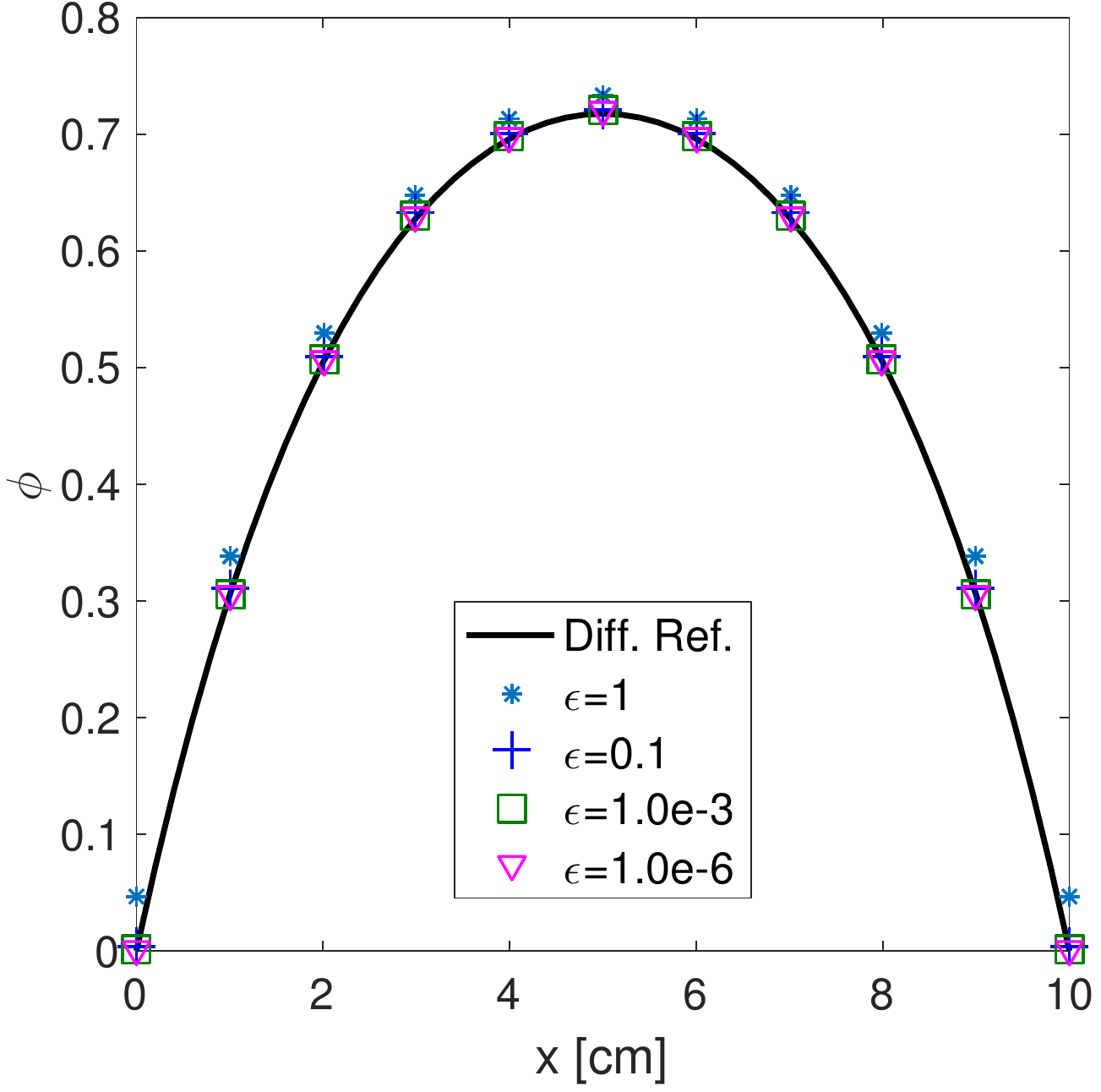}
	\caption{Asymptotic test for the thick diffusion limit using a uniform slab {with mesh size $h=1$\ cm}.}
	\label{f:asymp}
	
\end{figure}

The first test aims to numerically demonstrate the OT scaling preserves the thick diffusion limit as predicted in Section\ \ref{s:asym}.\ It is a 1D slab problem with a thickness of 10\ cm. Vacuum boundary conditions are imposed on both sides of the domain. The unscaled properties are: $\siga=0.01$\ \icm,\ $\st=5$\ \icm\ and $q=0.01$.\ Figure\ \ref{f:asymp}\ presents the OT-LS\pp{1}\ results with different $\epsilon$\ when performing the property scaling in Eq.\ \eqref{e:pscale}.\ A scheme preserving diffusion limit is expected to present a solution that approaches the diffusion solution as $\epsilon \rightarrow 0$.

\begin{figure}[ht!]
	
	\centering
	\includegraphics[width=.75\linewidth]{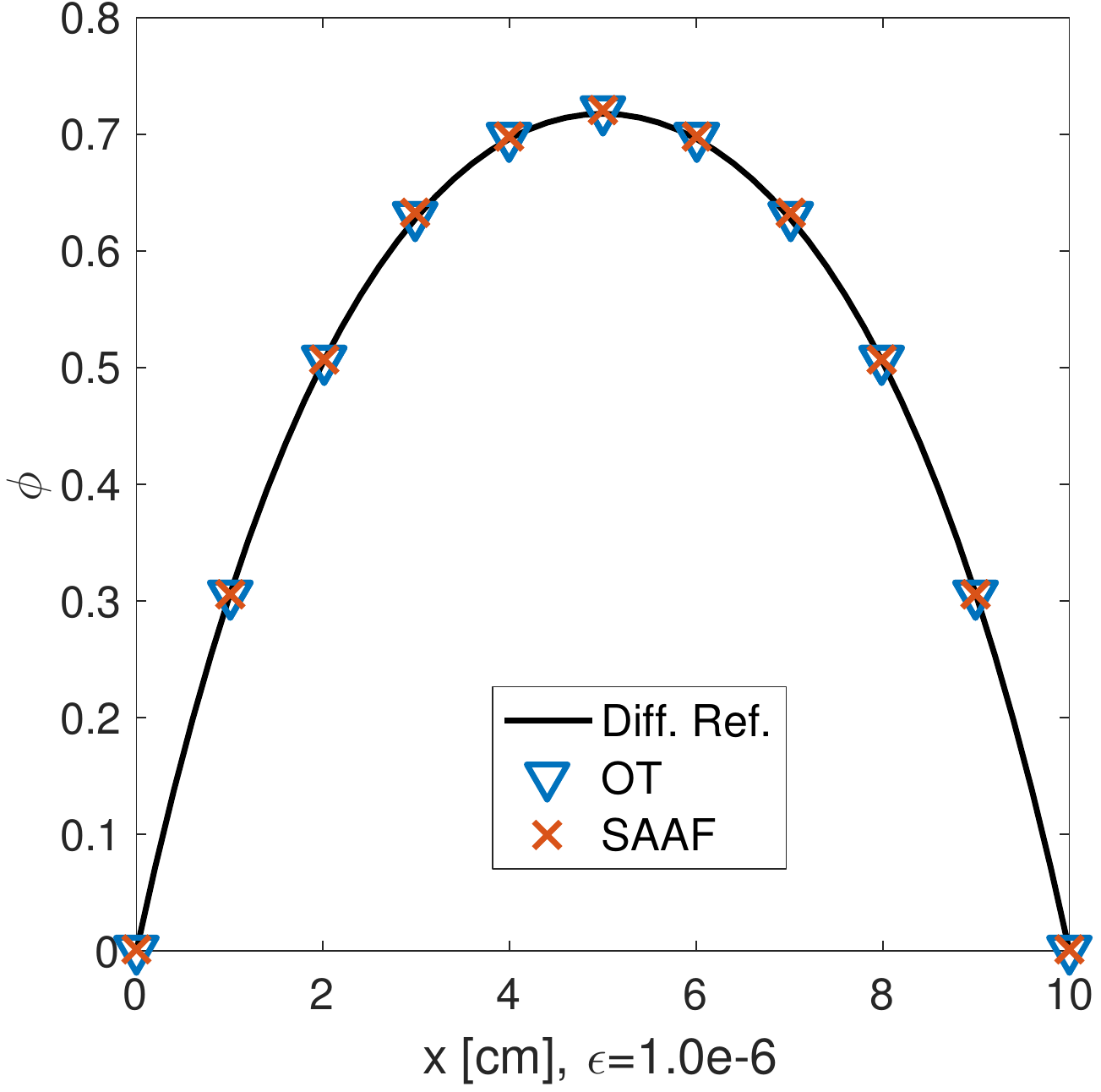}
	\caption{Comparison between SAAF and OT for $\epsilon=1.0\e{-6}$ on the first test problem {with mesh size $h=1$\ cm}.}
	\label{f:asymp2}
	
\end{figure}

Figure\ \ref{f:asymp}\ presents results for OT-LS\pp{1}\ with different $\epsilon$\ using 10 cells.\ When $\epsilon=1$,\ with which material properties are not scaled yet, the OT-LS\pp{1}\ presents slightly different solution to the diffusion, which is reasonable since angular discretization between OT-LS\pp{1}\ and diffusion are different. When decreasing $\epsilon$,\ however, OT-LS\pp{1}\ agrees with diffusion accurately. For other orders of \pn\ the results are similar and thus not shown. We also present the comparison between SAAF-\pn\ and OT-LS\pp{1}\ as shown in Figure\ \ref{f:asymp2}. Since SAAF-\pp{1}\ preserves the thick diffusion limit, the solution agrees with diffusion as well as OT-LS\pp{n}.\ Therefore, the results demonstrate the OT-LS\pp{n}\ as well as OT-LS\pp{1}\ preserves the thick diffusion limit.

\subsection{Thick problem with ``thin" materials with high scattering}
\begin{figure}[ht!]
	\centering
	\hspace*{0cm}\includegraphics[width=.75\linewidth]{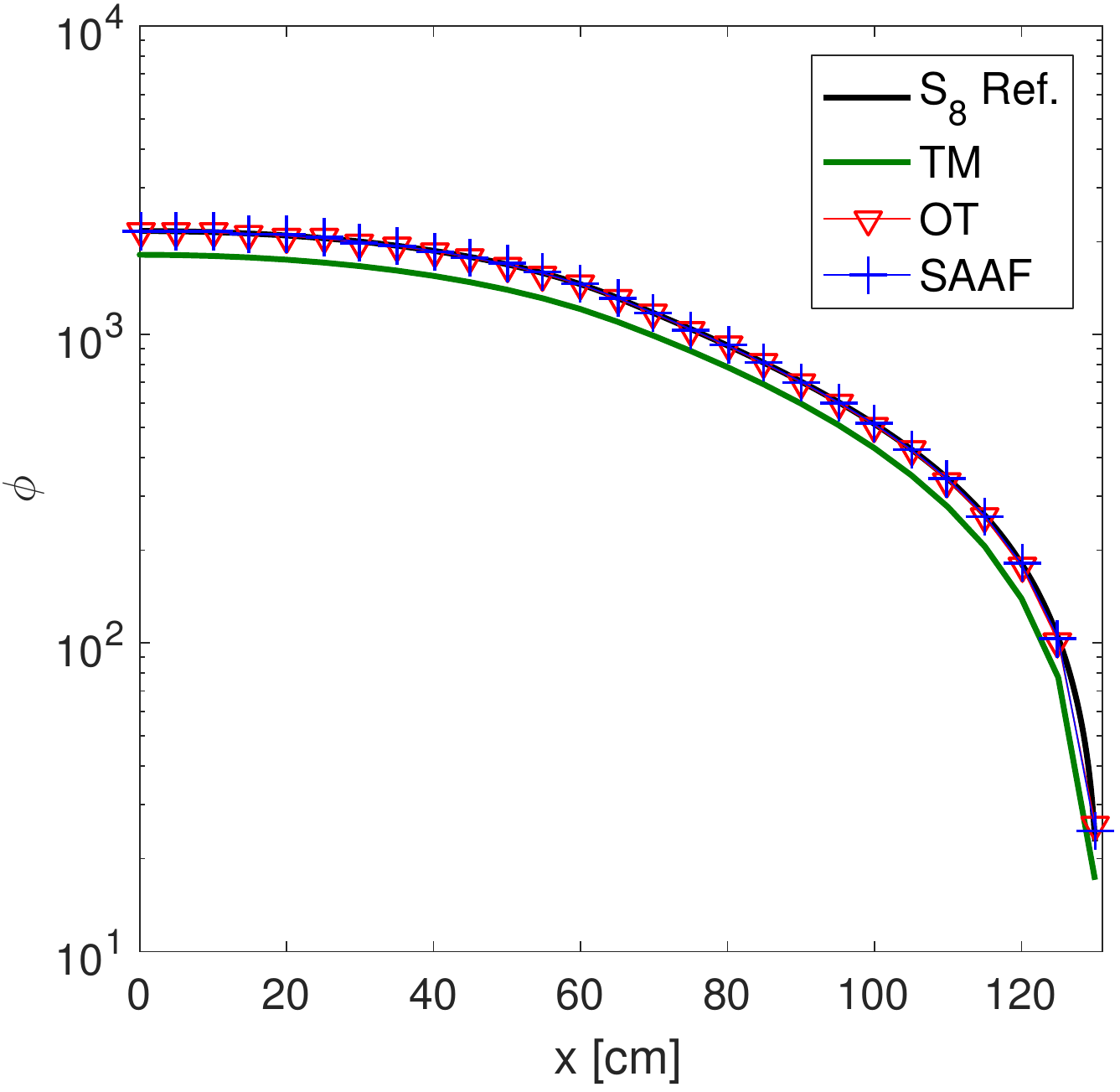}
	\caption{Natural carbon test with neutron at RT {with mesh size $h=5$\ cm.}}
	\label{bd1}
\end{figure}

\begin{figure}[ht!]
	\centering
	\hspace*{0cm}\includegraphics[width=.75\linewidth]{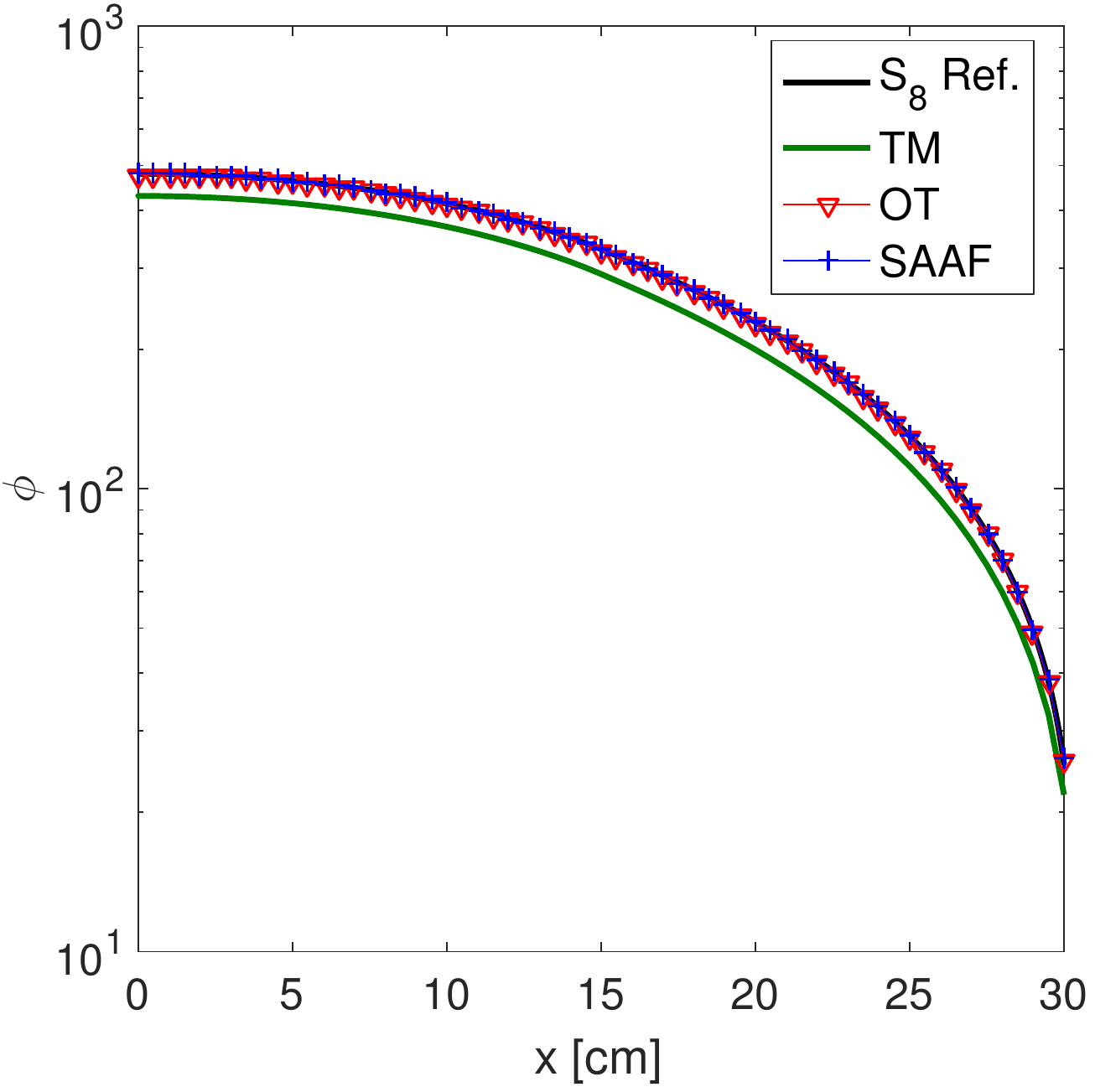}
	\caption{D$_2$O test with neutron at RT {with mesh size $h=0.5$\ cm}.}
	\label{bd2}
\end{figure}

Figure\ \ref{bd1}\ and \ref{bd2}\ {show scalar flux results for} scattering medium with moderate to large optical thickness. Figure\ \ref{bd1}\ shows the results with natural carbon throughout the domain for room-temperature (RT) neutrons ($\st=0.385$\ \icm\ and $\siga=3.2\e{-4}$\ \icm\ \cite{xie_book}). A unit source is imposed only in left half of the domain. A reflective boundary is imposed on left boundary. The overall thickness is {$50.05$}\ mean free paths. The reference solutions use  first-order S$_8$\ using diamond difference with 13000\ and 12000\ cells, respectively. Figure\ \ref{bd2}\ presents results with heavy water with similar settings as natural carbon test. The problem is {$13.47$}\ mean free paths ($\st=0.449$\ \icm\ and $\siga=3.3\e{-5}$\ \icm\ \cite{xie_book}). 

In both tests, the thick media scaling essentially imposes no scaling at all due to the fact that $\st<1.0$\ cm. The erroneous results indicate that even in problems with thin materials, scaling is necessary to preserve the solution accuracy. On the other hand, OT scaling presents comparable results to SAAF. 

Table \ref{t:t1} presents the $\tau_0$\ for SAAF, thick media, and OT scalings. On one hand, with unit scaling, the thick media scaling sacrifices accuracy. On the other hand, the SAAF has $\tau_0$\ over $10^3$\ or even $10^4$.\ In heterogeneous problems, this may induce several orders of scaling variations for different spatial regions, which might increase the risk of degrading linear solver performance and preconditioning efficiency. The OT scaling, however, makes a reasonable compromise that $\tau_0$\ is orders lower than SAAF-resembling scaling while reasonably preserving the accuracy in relatively thick scattering media.

\begin{table}[h]
	\centering
	\caption{$\tau_0$\ values for different scalings in carbon and heavy water tests.}
	\label{t:t1}
	\begin{tabular}{|c|c|c|c|}
		\hline
		Scalings/schemes&SAAF & TM & OT\\
		\hline
		$\tau_0$\ [cm]:\ carbon test&$3.125\e{3}$ & 1.0 & 125.125\\
		\hline
		$\tau_0$\ [cm]:\ D$_2$O test& $3\e{4}$ & 1.0 & 13.47\\
		\hline
	\end{tabular}
\end{table}
\subsection{1D iron water problem}

\begin{figure}[ht!]
	
	\centering
	\includegraphics[width=.75\linewidth]{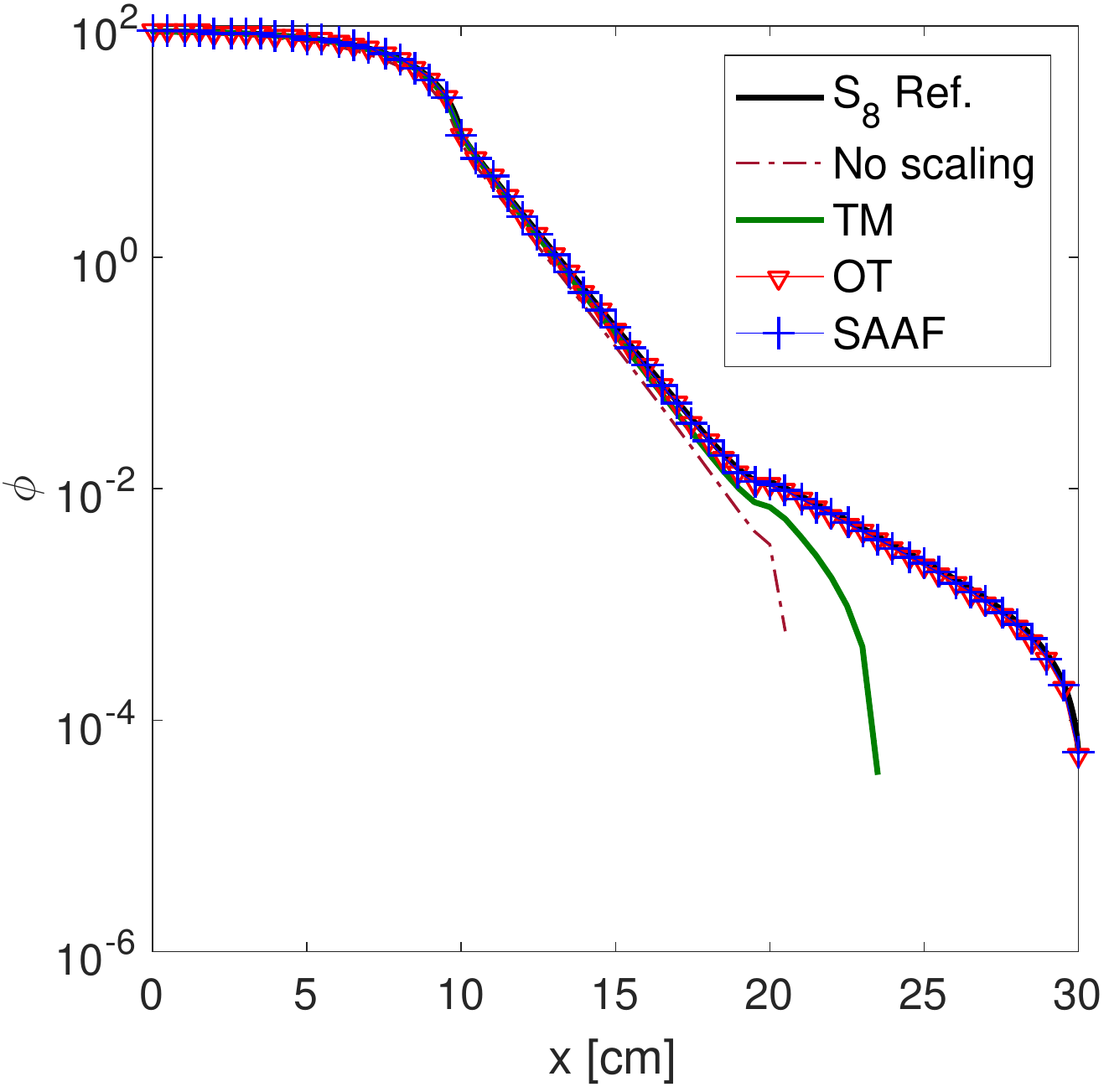}
	\caption{1D iron water problem with mesh size $h=0.5$\ cm.}
	\label{f:iw120}
	
\end{figure}

\begin{figure}[ht!]
	
	\centering
	\includegraphics[width=.75\linewidth]{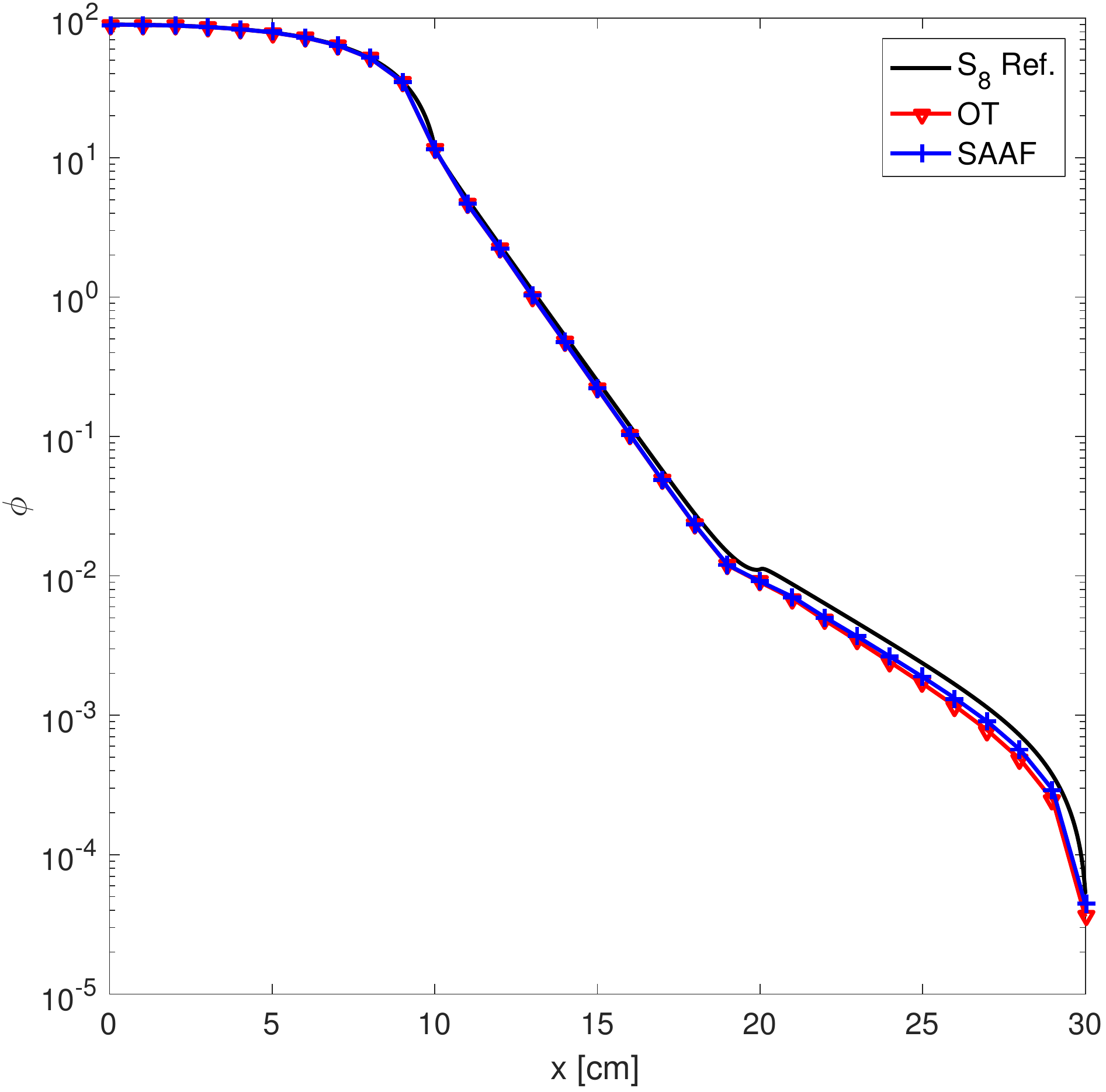}
	\caption{1D iron water problem with mesh size $h=1$\ cm.}
	\label{f:iw60}
	
\end{figure}

\begin{table}[ht!]
	\centering
	\caption{{Material cross sections in one-group iron-water test.}}
	\label{t:iw}
	\hspace*{-1cm}\begin{tabular}{|c|c|c|}
		\hline
		Materials & $\st$\ [cm$^{-1}$] & $\sigs$\ [cm$^{-1}$]\\
		\hline
		Water& $3.2759$ & $3.2656$ \\
		\hline
		Iron&$1.1228$&$0.9328$\\
		\hline
	\end{tabular}
\end{table}

For a final test, we use a modified iron-water shielding problem. The problem consists of a water source on the left side (10\ cm thick) with reflective boundary condition imposed on left, and a iron shield  (10\ cm thick) adjacent to the water source. Additionally, there is a layer of water on the right side of the problem  (10\ cm thick). The reference is generated with first-order S$_8$\ using 12000\ cells.

The material properties are from the thermal group in Ref.\ \cite{adams_iron_water}\ {and listed in Table \ref{t:iw}}.\ Both layers of water have an identical thickness of 32.759\ mean free paths. As seen in Figure\ \ref{f:iw120},\ unscaled LS\pp{7}\ presents a negative scalar flux in the right water shield. With the thick media scaling, the scalar flux is improved but still with the presence of negative answers. This problem is a deep penetration problem with nearly seven orders of change in scalar flux. The failure of these two methods demonstrate the necessity of constructing a proper scaling in this type of problems.

On the other hand, though $\tau_0$\ in OT differs from $1/\siga$\ in SAAF, both methods present comparably accurate solutions when the mesh is relatively fine as in Figure\ \ref{f:iw120}.\ With a coarser mesh, though solution accuracy decreases, especially in the right water shield, OT still presents a reasonably accurate solution in the other part of the domain.
\section{Conclusions and Future Work}
\label{s:con}
\subsection{Conclusions}
In this paper, we formulated unscaled and scaled least-squares discretizations in space combined with \pn\ in angle (LS\pn) for solving neutron transport problems.

We also developed a novel scaling based on optical thickness of scattering medium which preserves the thick diffusion limit as demonstrated by an asymptotic test with continuous finite element basis functions. The scaling is such that in problems with tens to hundreds of mean free paths with highly scattering mediums such as heavy water and carbon the zeroth moment equation in LS\pn\ is scaled by a factor proportional to the optical thickness. Otherwise, reciprocal removal cross sections are used. The scaling improves Manteuffel, et al.'s TM scaling in highly scattering media. Also, the boundedness of the scaling will avoid the solving and preconditioning degradation which occurs to SAAF-\pn\ in multi-D applications in highly scattering mediums. In all the tests presented in this paper, the scaling is accurate and robust.

\subsection{Future work}
{We considered isotropic scattering in this work. However, the method effectiveness in cases where anisotropic scattering manifests is not known. Further investigations, such as appropriate modifications of the scaling for anisotropic scattering, are recommended for future work.}

{In addition, we developed the optical-thickness scaling for LS\pn\ in this work. However, it can be realized that extending the scaling to LS\sn\ is minor complication and would be an interesting future direction.}

{Last but not least, extending the method to multi-D is worth studying. Our model requires $D$\ and $L$\ to calculate the scaling parameter $\tau_0$\ in Eq.\ \eqref{e:tau}.\ Though it is straightforward to provide $D$\ and $L$\ in 1D slab geometry, systematic methods of determining these two parameters in multi-D geometry need further investigations, especially for materials with complex shapes using non-uniform meshes. It is likely that using characteristic lengths in these geometries could be sufficient.}


%
\section*{ACKNOWLEDGMENTS}
This project is funded by Department of Energy NEUP research grant from Battelle Energy Alliance, LLC- Idaho National Laboratory, Contract No: C12-00281.

W.\ Zheng would thank Dr.\ Vincent M. Laboure for fruitful discussions regarding a draft of this manuscript. The authors wish to thank the anonymous reviewers with their invaluable advice to improve the quality of this paper.
\section*{References}
\bibliography{mybibfile}

\end{document}